\documentclass[twocolumn,pra,showpacs,amsmath,amssymb]{revtex4}

\usepackage{amsmath,amsfonts,amsthm,graphicx,color}

\usepackage{float}
\usepackage{hyperref}

\begin{document}

\title{Lower bound on the communication cost of simulating bipartite quantum correlations}
\author{T. V\'ertesi}
\email{tvertesi@dtp.atomki.hu}
\affiliation{Institute of Nuclear Research of the Hungarian Academy of Sciences\\
H-4001 Debrecen, P.O. Box 51, Hungary}
\author{E. Bene}
\email{bene@atomki.hu}
\affiliation{Institute of Nuclear Research of the Hungarian Academy of Sciences\\
H-4001 Debrecen, P.O. Box 51, Hungary}

\def\CC{\mathbb{C}}
\def\RR{\mathbb{R}}
\def\one{\leavevmode\hbox{\small1\normalsize\kern-.33em1}}
\newcommand*{\tr}{\mathsf{Tr}}
\newcommand*{\sgn}{sgn}
\newtheorem{theorem}{Theorem}[section]
\newtheorem{lemma}[theorem]{Lemma}

\date{\today}

\begin{abstract}

Suppose Alice and Bob share a maximally entangled state of any
finite dimension and each perform two-outcome measurements on the
respective part of the state. It is known, due to the recent result of Regev and Toner,
that if a classical model is augmented with two bits of communication then all the
quantum correlations arising from these measurements can be
reproduced. Here we show that two bits of communication is in fact
necessary for the perfect simulation. In particular, we prove that
a pair of maximally entangled four-dimensional quantum systems
cannot be simulated by a classical model augmented by only one bit
of communication.

\end{abstract}

\maketitle

\section{Introduction}\label{intro}

Let us assume two spatially separated parties, Alice and Bob, who
receive local inputs and produce subsequently outputs. As pointed
out by John Bell \cite{Bell64}, joint correlations of Alice's and
Bob's outputs resulting from quantum theory cannot be reproduced
classically if communication is not taken place between them.
Actually, his proof consists a setup of a gedanken experiment
where contradiction arises if only local resources are used by the
two parties. This is exemplified by the violation of the so-called
CHSH inequalities \cite{CHSH69}, which has been demonstrated
experimentally several times up to some technical loopholes
\cite{Aspect}. Considering that local resources are not sufficient
to simulate all the correlations of quantum mechanics one may ask
whether Alice and Bob would benefit from sending some bits of
communication in order to reproduce all the set of quantum
correlations. This problem has been addressed by Maudlin
\cite{Maudlin} and later independently by Brassard et al.
\cite{Brassard99} and Steiner \cite{Steiner00}. However, the
approach they used in defining the communication cost of the
simulation is slightly different. Brassard et al. took the worst
case communication, i.e., the maximal number of bits to be sent in
the worst case to simulate quantum correlations, while in
Steiner's and Maudlin's model the average amount of communication
is considered. In this work we apply the worst case scenario,
however, there are several works discussing the cost of average
communication for simulation (see for example,
\cite{Cerf00,Pironio03,Methot04}). In this respect we also wish to
highlight two more recent works. One is by Gavinsky
\cite{Gavinsky}, who present a nonlocality game, which requires a
large amount of communication so as to be simulated classically.
In an other work by Roland and Szegedy \cite{RS}, an asymptotic lower
bound is established on the cost of simulating quantum
correlations.

In Ref.~\cite{Brassard99} the intriguing result has been found
that one can simulate all quantum correlations arisen from
two-outcome projective measurements on the two-qubit maximally
entangled state, provided a finite (eight) number of bits of
communication is added to local resources. This result has been
slightly improved later by Csirik to six bits \cite{Csirik02}.
Then six bits could be further improved even to one bit by the
protocol of Toner and Bacon \cite{TB03} in 2003. One may assume
that this result is due to the fact that the Hilbert space of each
party is restricted to two dimensions. So the question was natural
to ask that how many bits are needed to simulate correlations
arising from two-outcome projective measurements without
restricting the size of the Hilbert space. First, it has been
exhibited a protocol by Toner and Regev (see \cite{TonerPhD})
which solves this problem in any Hilbert space dimension with five
bits, moreover this result was subsequently improved to two bits
\cite{RT07}. This is the best upper bound up to now, and an open
question is whether two bits of communication are needed at all,
or the minimal amount of one bit is enough to simulate all
two-outcome projective measurements on a maximally entangled state
of any finite dimension. We wish to address this problem and show
that two bits are indeed necessary, as conjectured by Regev and
Toner \cite{RT07}, by exhibiting a pair of four-dimensional
quantum systems and measurement settings on Alice's and Bob's part
so that the corresponding quantum correlations cannot be simulated
with a classical model using one bit of communication. Note, that
for more parties even the question of finite communication is not
settled yet. In this respect, Broadbent et al. \cite{BCT08} proved
that at least $n\log n-2n$ bits are needed to simulate classically
an $n$-party GHZ state improving on an earlier result of Buhrman
et al. \cite{BHMR03}.

In the two-party two-outcome case, on the other hand, Bacon and
Toner \cite{BT03} have shown that for three measurements per party
one bit of communication suffices to solve this task. We will show
in this work that for an infinite number of measurements the
exchange of one bit is not sufficient. We believe that this result
holds for finite number of settings as well, but we leave it as an
open problem. Note that to the best of our knowledge there is no
two-party protocol in the literature even for the general
$d\ge2$-outcome case where it is proved that quantum measurements
cannot be simulated classically with one bit of information.

In the quantum two-party two-outcome scenario Alice and Bob share
an entangled state $\rho$ and on the respective parts they perform
projective measurements described by the observables
$\mathbf{A}(a)$ and $\mathbf{B}(b)$ with eigenvalues $\pm 1$ for
Alice and Bob, respectively. Here $a$ and $b$ label the
observables. Then Alice and Bob each output two bits, $\alpha$ and
$\beta$. In the quantum setting the correlation
$E[\alpha\beta|ab]$ satisfies $E[\alpha\beta|ab]=\tr(
\mathbf{A}(a)\otimes \mathbf{B}(b)\rho)$.

As Tsirelson has shown, there is an equivalent formulation of this
problem by means of unit vectors \cite{Tsirelson}: Alice receives
as an input the unit vector $a\in R^n$ and outputs a bit
$\alpha\in\pm 1$, also Bob receives as an input the unit vector
$b\in R^n$ and outputs a bit $\beta\in\pm 1$. Their goal is to
produce a correlation $E[\alpha\beta|ab]$ which satisfies
$E[\alpha\beta|ab]=A(a)\cdot B(b)$, where $A(a)$ and $B(b)$ are some
unit vectors in $R^m$ and $\cdot$ denotes dot product.
Then, due to Tsirelson's construction \cite{Tsirelson},
these correlations can be realized as $\pm 1$-valued observables
$\mathbf{A}(a)$ and $\mathbf{B}(b)$ on a maximally entangled state
of local dimension $D=2^{\lfloor m/2\rfloor}$.

Such as in the context of Bell inequalities, let us now take a
linear function of the correlations
\begin{equation}
B(M)=\int_{a,b\in S^{n-1}}{M(a,b)A(a)\cdot B(b)d\sigma(a) d\sigma
(b)},\label{BM}
\end{equation}
where we have the functions $A,B: S^{n-1} \to S^{m-1}$. Thus
$B(M)$ can be interpreted as a two-party two-outcome Bell
expression with a continuous number of measurement settings. Let
us fix $M(a,b)=a\cdot b$, which is the famous example studied by
Grothendieck \cite{Grothendieck}, and was also a subject of recent
studies \cite{Briet,VP09} related to dimension witnesses
\cite{Brunner08}. We wish to find the following values:
\begin{align}
Q(n)&=\max_{A,B: S^{n-1} \to S^{m-1}}{B(M)},\label{Qn}\\
L(n)&=\max_{A,B: S^{n-1} \to \pm1}{B(M)}, \label{Ln}
\end{align}
where $Q(n)$ is the maximum value of the Bell expression $B(M)$
achievable by means of quantum systems (without imposing bound on
the value of $m$), while $L(n)$ corresponds to the maximum $B(M)$
attainable by local hidden variables systems.

In Sec.~\ref{Q} we give a lower bound to $Q(n)$ by setting
particularly $A(a)=a$ and $B(b)=b$, which bound is known to be the
exact maximum value as well (see \cite{Briet},\cite{VP09}). Then,
in Sec.~\ref{L} the local bound $L$ is given, which was also
considered by Grothendieck himself. Next, in Sec.~\ref{C} we
evaluate the local bound augmented with one bit of communication,
which we denote by $C(n)$. Since $M(a,b)=a\cdot b$ is symmetric in
the vectors $a$ and $b$, it is enough to treat the case where
Alice may transmit one bit of information to Bob. In this case Bob's
outcome may depend on the input Alice received. Thus, by allowing
one bit of communication, the $\pm 1$ valued function $B(b)$ of a
locally classical model becomes $B(b,f(a))$. However, the function
$f(a)$ may depend only on a bipartition of the set $a$, since a
bipartition carries just one bit of information. Then substituting
$B(b,f(a))$ in place of $B(b)$ in Eq.~(\ref{BM}), one gets
\begin{align}
&C(S',S'',n)=\max_{\substack{A:S^{n-1}\to\pm 1\\B':S'\to\pm 1}}\int{(a\cdot b)A(a)\cdot B'(b)d\sigma(a)d\sigma(b)}\nonumber \\
&+\max_{\substack{A:S^{n-1}\to\pm 1\\B'':S''\to\pm 1}}\int{(a\cdot b)A(a)\cdot
B''(b)d\sigma(a)d\sigma(b)}, \label{CSn}
\end{align}
where $S^{n-1}=S'\cup S'$. Then the maximum value achievable by a
local model plus one bit of communication is
\begin{equation}
C(n)=\max_{S'/S''}C(S',S'',n),\label{Cn}
\end{equation}
that is, we have to maximize with respect to all possible
bi-partitions of the unit sphere $S^{n-1}$.

The main result of this paper is the proof that
$C(n)<\tilde{Q}(n)$ for $n\ge 5$, where $\tilde{Q}$ is a lower
bound to $Q(n)$ (however, due to \cite{Briet,VP09} this bound is
tight). This implies by the construction of Tsirelson, that
measurements on maximally entangled four-dimensional systems
(ququarts) cannot be simulated by a local classical model allowing
only one bit of communication.

\section{Calculation of the limits}\label{limits}

\subsection{Quantum bound}\label{Q}

A lower bound to the value of $Q(n)$ can be given by substituting
$A(a)=a$ and $B(b)=b$ into the definition~(\ref{Qn}), where the
Bell expression $B(M)$ is defined by (\ref{BM}). Then $m=n$ and a
lower bound to $Q(n)$ is given by
\begin{equation}
\tilde{Q}(n)=\int_{a,b\in S^{n-1}}{|a\cdot b|^2
d\sigma(a)d\sigma(b)}\le Q(n), \label{Qn2}
\end{equation}
where $\int{d\sigma(a)}=\int{d\sigma(b)}=1$ is the normalized Haar
measure. Due to rotational invariance we can assume
$b=(1,0,...,0)$, thus we can further write (\ref{Qn2}) to obtain
$\tilde{Q}(n)=\int{|b_1|^2
d\sigma(b)}=(1/n)\int\sum_i{|b_i|^2}d\sigma(b)=1/n$, where we used
that all $b_i$, $i=1,\ldots,n$ are equal owing to symmetry
arguments. As it has been shown \cite{Briet,VP09}, $\tilde{Q}(n)$
is equal to $Q(n)$, but in our proof we do not have to use this
fact.

\subsection{Local bound}\label{L}

Here we evaluate the classical limit $L(n)$. According to
(\ref{Ln}) and using the explicit form $M(a,b)=a\cdot b$, this is
\begin{equation}
L(n)=\max_{A,B:S^{n-1}\to\pm 1}{\int{(a\cdot b)
 A(a)B(b)d\sigma(a)d\sigma(b)}},\label{Ln2}
\end{equation}
where now $A(a)$ and $B(b)$ are $\pm 1$ valued functions.

Let us define $h(a)=\int{(a\cdot b) B(b)d\sigma(b)}= a\cdot \int{b
B(b)d\sigma(b)} = a\cdot Z = \lambda a\cdot z$, where $z$ is a
unit vector and $\lambda=h(z)$ is the length of the vector $Z$.
Therefore, $h(a)=h(z) a\cdot z$ and according to (\ref{Ln2})
\begin{align}
L(n)&=\max_{A,B:S^{n-1}\to\pm 1}{\int{A(a)h(a)d\sigma(a)}}\nonumber\\
&= \max_{A,B:S^{n-1}\to\pm 1}h(z)\int{(z\cdot
a)A(a)d\sigma(a)},\label{Ln3}
\end{align}
where $z$ in the second line is defined through
\begin{equation}
Z=\lambda z=\int_{b\in S^{n-1}}{b B(b)d\sigma(b)} \label{Z}
\end{equation} according to the formulae below Eq.~(\ref{Ln2}).
Then (\ref{Ln3}) above can be further written as
\begin{align}
&L(n)=\max_{A,B:S^{n-1}\to\pm 1}{\int{(z\cdot b)B(b)d\sigma
(b)}\int{(z\cdot a) A(a)d\sigma(a)}}\nonumber\\
&=\left[\int_{a\in S^{n-1}}{|z\cdot
a|d\sigma(a)}\right]^2,\label{Ln4}
\end{align}
where the second line follows from the substitutions
$A(a)=\sgn(z\cdot a)$ and $B(b)=\sgn(z\cdot b)$, which choices are
incidentally consistent with the definition~(\ref{Z}) for $z$. Due
to symmetry arguments one can choose $z=(1,0,\ldots,0)$ and then
we have $L(n)=\left[\int|a_1|d\sigma(a)\right]^2$. By an explicit
calculation one obtains $L(n)=s_{n-1}/(n s_n)$ (see also
\cite{Briet,VP09}), where
$s_n=\int_0^{\pi}{\sin^n\vartheta}d\vartheta=\sqrt\pi\Gamma((n+1)/2)/\Gamma((n+2)/2)$.
This formula can be evaluated analytically either for small values
of $n$ or for the continuum limit $n\rightarrow\infty$. We have
collected the ratios $\tilde{Q}(n)/L(n)$ for some small values of
$n$ and also for $n\rightarrow\infty$ in Table~\ref{table1}
(remembering that $\tilde{Q}(n)=1/n$ for $n>1$).

\begin{table*}[tbm]
 \caption{The lower bound $\tilde{Q}(n)/L(n)$ by the values of $n=2,3,4,5,6$ and $n\rightarrow\infty$.}
 \vskip 0.2truecm
 \centering
 \begin{tabular}{l c c c c c c }
 \hline\hline
 n&2&3&4&5&6&$n\rightarrow\infty$\\ [0.5ex]
 \hline\\ [-2ex]
 $\frac{\tilde{Q}_n}{L_n}$\hphantom{ *}&$\frac{\pi^2}{8}\sim 1.234$\hphantom{ *}&$\frac{4}{3}\sim 1.333$\hphantom{ *}&$\frac{9\pi^2}{64}\sim 1.388$\hphantom{ *}&$\frac{64}{45}\sim 1.422$\hphantom{ *}&
 $\frac{75\pi^2}{517}\sim1.432$\hphantom{ *}&$\frac{\pi}{2}\sim 1.577$\hphantom{
 *}\\ [1ex]
 \hline
 \end{tabular}
 \label{table1}
 \end{table*}

\subsection{Local plus one bit bound}\label{C}

Let us now calculate an upper bound on the value of $C(n)$ defined
by Eq~(\ref{Cn}). We use the same argumentations to separate the
terms involving the vectors $a$ and $b$, which lead from
(\ref{Ln2}) to (\ref{Ln4}). Then, the first double integral in
(\ref{CSn}) can be upper bounded as
\begin{align}
&\max_{\substack{A:S'\to\pm 1\\B':S^{n-1}\to\pm 1}}\int{(a\cdot b)
A(a)B'(b)d\sigma(a)d\sigma(b)}\le\max_{z'}\nonumber\\&\left\{\max_{A:S'\to\pm
1}\int{(z'\cdot a) A(a)d\sigma(a)}\max_{B':S^{n-1}\to\pm
1}\int{(z'\cdot b) B'(b)d\sigma(b)}\right\}\nonumber\\
&= \max_{z'}\left\{\int_{a\in S'}{|z'\cdot a|d\sigma(a)}\int_{b\in
S^{n-1}}{|z'\cdot b|d\sigma(b)}\right\}, \label{ineq}
\end{align}
where $z'$ is defined by the normalized value of $Z'=\int_{b\in
S^{n-1}}{b B'(b)}d\sigma(b)$. The second term in (\ref{CSn}) can
be written similarly to (\ref{ineq}) but replacing the terms $(')$
by $('')$. In order to arrive at the third line we used the
substitutions $A(a)=\sgn(z'\cdot a)$ and $B'(b)=\sgn(z'\cdot b)$,
$a\in S',b\in S^{n-1}$ (and likewise for case $('')$ we will have
$A(a)=\sgn(z''\cdot a)$ and $B''(b)=\sgn(z''\cdot b)$, $a\in S'',b
\in S^{n-1}$). By the virtue of the definition of $Z'$ (and $Z''$)
above, for any alignment of the vectors $z'$ and $z''$, the
functions $B'$ and $B''$ are defined consistently. On the other
hand, due to rotational invariance, the last integral term in the
third line of (\ref{ineq}) does not depend on the vector $z'$. By
exploiting this fact also in the expression $('')$ analog
with~(\ref{ineq}) and by using the the definitions in
(\ref{CSn},\ref{Cn}), we can write
\begin{align}
&C(n)\le\int_{b\in S^{n-1}}{|b_1|d\sigma(b)}\nonumber\\
&\max_{z',z''}\max_{S'/S''}\left\{\int_{a\in S'}{|z'\cdot
a|d\sigma(a)}+\int_{a\in S''}{|z''\cdot
a|d\sigma(a)}\right\}\nonumber\\&=\int_{b\in
S^{n-1}}{|b_1|d\sigma(b)}\max_{z',z''}\int_{a\in
S^{n-1}}{\max\left\{|z'\cdot a|,|z''\cdot a|\right\}d\sigma(a)}. \label{Cn2}
\end{align}
Since only the angle between $z'$ and $z''$ enters above, without
loss of generality we can choose them as
\begin{align}
z'&=(\sin\theta,\cos\theta,0,\ldots,0),\nonumber\\
z''&=(-\sin\theta,\cos\theta,0,\ldots,0). \label{z12}
\end{align}
Now, taking into account Eq.~(\ref{Ln4}) we have
\begin{equation}
C(n)\le\sqrt{L(n)}\max_{z',z''}\int_{a\in S^{n-1}}{\max\left\{|z'\cdot
a|,|z''\cdot a|\right\}d\sigma(a)}, \label{int}
\end{equation}
which readily depends only on the angle $2\theta$ between $z'$ and
$z''$ in (\ref{z12}). Also, the RHS of~(\ref{int}) above is
invariant under the sign changes of $z', z''$ and $a$. Thus, it is
enough to consider the angle $0\le \theta\le \pi/2$ and $a$ being
located in the first quadrant of the first two coordinates.
However, in this case $|z'\cdot a|>|z''\cdot a|$, and the integral
in~(\ref{int}) becomes $4\max_{\theta}{\int_{a\in
S^{n-1}_*}|z'\cdot a|d\sigma(a)}$, where the integral is performed
on the sphere $S^{n-1}$ except in the first two coordinates, where
integration is only over the first quadrant (the range of $a$
designated by $S^{n-1}_*$).

Let us next evaluate an upper bound on the ratio $C(n)/L(n)$,
i.e., on the ratio of the local bound with one bit of
communication to the local bound without communication. Applying
Eqs.~(\ref{Cn}), (\ref{Ln4}) and (\ref{int}), we have
\begin{align}
\frac{C(n)}{L(n)}&\le\frac{4\max_{\theta}{\int_{a\in
S^{n-1}_*}|z'\cdot
a|d\sigma(a)}}{\int_{a\in S^{n-1}}|a_1|d\sigma(a)}\nonumber\\
&=\frac{\max_{\theta}\int_{\vartheta=0}^{\pi/2}{\sin(\theta+\vartheta)d\vartheta}}
{\int_{\vartheta=0}^{\pi/2}{\sin(\vartheta)d\vartheta}}\nonumber\\
&=\max_{\theta}\{\cos\theta + \sin\theta\}=\sqrt 2\sim 1.41421,
\end{align}
the maximum taken up by $\theta=\pi/4$, where we have taken in the
integration the explicit forms of $z'$ and $z''$ from (\ref{z12}).
Notice, that this upper bound to the ratio is independent of
dimension $n$. However, a lower bound on the quantum per local
bound can be seen in Table~\ref{table1}. This shows clearly that
for $n=5$ the ratio exceeds the value $\sqrt 2$, indicating that
quantum mechanical correlations of bipartite quantum systems
cannot be simulated by local models augmented with one bit of
communication.

Due to the work of Tsirelson, for $n=5$ one can construct the measurement
operators and states in the local four-dimensional Hilbert spaces.
In the following, we exhibit the explicit form of them. Let the
quantum state be the maximally entangled pair of ququarts,
$|\psi^+\rangle=(1/2)\sum_{i=1}^4{|ii\rangle}$. Then, the
respective observables of Alice and Bob are
\begin{align}
\mathbf{A}(a)&=\sum_{k=1}^5{A^{(k)}(a)\gamma_k}\nonumber\\
\mathbf{B}(b)&=\sum_{k=1}^5{B^{(k)}(b)\gamma_k},
\end{align}
where $A^{(k)}(a),B^{(k)}(b)$ are the components of the vectors
$A(a)$ and $B(b)$ and the five anticommuting, traceless $\gamma$
matrices are
\begin{align}
\gamma_1&=\sigma_x\otimes\one\nonumber\\
\gamma_2&=\sigma_y\otimes\one\nonumber\\
\gamma_3&=\sigma_z\otimes\sigma_x\nonumber\\
\gamma_4&=\sigma_z\otimes\sigma_y\nonumber\\
\gamma_5&=\sigma_z\otimes\sigma_z.
\end{align}
It can be checked that with this, one has indeed $\langle
\psi^+|A(a)\otimes B(b)|\psi^+\rangle=A(a)\cdot B(b)$ as
required.

\section{Summary}\label{sum}

In this paper we have shown that two bits of communication are
necessary for perfectly simulating classically the correlations of
measurement outcomes carried out by two distant parties. In
particular, we proved that two-outcome projective measurements on
a pair of maximally entangled four-dimensional quantum systems
cannot be simulated by a classical model augmented by only one bit
of communication. In order to prove it, our scenario involved an
infinite number of measurements. We pose it as an open question
whether a finite number of measurements would suffice for the
proof as well. In this respect we mention that in a recent work
Bri\"{e}t et al. \cite{Briet} could discretize the Bell expression
(\ref{BM}) with $M=a\cdot b$ to involve only finite number of
measurement settings and obtained bounds on the maximum quantum
values depending on the dimension. Their result would probably
help in the one bit communication problem, discussed here, as
well.

Finally, we list some interesting open questions related to recent
results in the literature. Let us restrict to the case of two
parties and binary outputs. We know that in this case measurements
can be simulated classically with one bit (two bits) of
communication performed on maximally entangled qubits \cite{BT03}
(qudits \cite{RT07}). Recently, N.~Gisin posed the question
\cite{Gisin07}, whether there exist measurements on a pair of
partial entangled qubit states which cannot be simulated by a
single bit of communication. Similarly, it would be interesting to
know how hard to simulate partially entangled qudits, e.g. whether
it would require exchanging more than two bits. Also, it is known
\cite{CGMP05} that a hypothetical non-local machine, the so-called
PR-box \cite{PR} is a strictly weaker resource than one bit of
communication. Nevertheless, it can simulate two-outcome
projective measurements on a maximally entangled qubit pair
\cite{CGMP05}. On the other hand, it has been recently proved that
projective measurements on a maximally entangled pair of qudits
with two outcomes can be simulated by three PR-boxes
\cite{KKLR09}. In light of our result that no 1-bit communication
model exists for simulating measurements on maximally entangled
qudits, no 1 PR-box model would exist either. Then it would be
interesting to find out whether 2 PR-boxes were enough to simulate
two-outcome projective measurements on maximally entangled qudits or not.

\acknowledgments T.V. has been supported by a J\'anos Bolyai
Programme of the Hungarian Academy of Sciences. We wish to thank
Ben Toner, Oded Regev and an anonymus Referee for filling gaps 
in the proof of Sec.~IIC. We also thank Nicolas Brunner for pointing 
out the lower bound result on PR-boxes.

\end{document}